\documentclass[useAMS]{mn2e}
\usepackage{psfig}
\newif\ifAMStwofonts

\def\be{\begin{eqnarray}}
\def\ee{\end{eqnarray}}
\def\beq{\begin{equation}}
\def\eeq{\end{equation}}

\def\etal{{\it et al.}}

\def\HI{\hbox{H~$\scriptstyle\rm I\ $}}
\def\HII{\hbox{H~$\scriptstyle\rm II\ $}}
\def\HeI{\hbox{He~$\scriptstyle\rm I\ $}}
\def\HeII{\hbox{He~$\scriptstyle\rm II\ $}}
\def\HeIII{\hbox{He~$\scriptstyle\rm III\ $}}
\def\SiIV{\hbox{Si~$\scriptstyle\rm IV\ $}}
\def\CIV{\hbox{C~$\scriptstyle\rm IV\ $}}
\def\nHI{{\rm HI}}
\def\nHII{{\rm HII}}
\def\nHe{{\rm He}}
\def\nHeI{{\rm HeI}}
\def\nHeII{{\rm HeII}}

\def\lya{Ly${\alpha}\,\,$}

\def\lesssim{\mathrel{\hbox{\rlap{\hbox{\lower4pt\hbox{$\sim$}}}\hbox{$<$}}}}
\def\gtrsim{\mathrel{\hbox{\rlap{\hbox{\lower4pt\hbox{$\sim$}}}\hbox{$>$}}}}

\def\gtsima{$\; \buildrel \over \sim \;$}
\def\ltsima{$\; \buildrel < \over \sim \;$}
\def\prosima{$\; \buildrel \propto \over \sim \;$}
\def\gsim{\lower.5ex\hbox{\gtsima}}
\def\lsim{\lower.5ex\hbox{\ltsima}}
\def\simgt{\lower.5ex\hbox{\gtsima}}
\def\simlt{\lower.5ex\hbox{\ltsima}}

\def\simpr{\lower.5ex\hbox{\prosima}}
\def\la{\lsim}

\def\etal{{\frenchspacing etal. }}
\def\ie{{\frenchspacing\it i.e. }}
\def\eg{{\frenchspacing\it e.g. }}

\def\be{\begin{eqnarray}}
\def\ee{\end{eqnarray}}

\def\CR{{\tt CRASH }}

\def\Lya{Ly$\alpha$ }

\title[Radiative Transfer Effects on the \lya Forest]{Radiative Transfer Effects on the \lya Forest}

\author[A. Maselli \& A. Ferrara]{
A. Maselli$^{1}$  and A. Ferrara$^{2}$ \\
$^1$ Dipartimento di Astronomia, Universit\'a di Firenze, Largo
Enrico Fermi 5, 50125 Firenze, Italy\\
$^2$ SISSA/International School for Advanced Studies, via Beirut
2-4, 34014 Trieste, Italy\\ }

\date{\today}
\pagerange{\pageref{firstpage}--\pageref{lastpage}} \pubyear{2004}
\begin{document}

\maketitle
\label{firstpage}

\begin{abstract}
Strong observational evidence for a fluctuating
ultraviolet background (UVB) has been accumulating through
a number of studies of the \HI and \HeII \Lya forest as well as
accurate IGM metallicity measurements. UVB fluctuations could
arise both from the inhomogeneous distribution of the ionizing
sources and/or from radiative transfer (RT) through the filamentary
IGM. In this study we investigate, via numerical simulations, the
role of RT effects such as shadowing, self-shielding and filtering
of the ionizing radiation, in giving raise to a fluctuating UVB. We
focus on possible detectable signatures of these effects on
quantities derived from \lya forest spectra, as photoionization
rate fluctuations, $\eta$($\equiv N_\nHeII/N_\nHI$) parameter distributions 
and the IGM temperature at $z\approx 3$. We find
that RT induces fluctuations up to 60\% 
in the UVB, which are tightly correlated to the density field.
The UVB mean intensity is progressively suppressed toward
higher densities and photon energies above 4 Ryd, 
due to the high \HeII opacity. 
Shielding of overdense regions ($\Delta \gtrsim
5$) from cosmic \HeII ionizing radiation, produces a decreasing trend
of $\eta$ with overdensity. 
Furthermore we find that the mean $\eta$ value 
inferred from \HI-\HeII \lya forest observations can be explained only
by properly accounting for the actual IGM opacity. 
We outline and discuss several  implications of our findings.
\end{abstract}


\section{Introduction}

The lack of the Gunn-Peterson trough in the spectra of quasars at redshift $z < 5$ indicates that the 
intergalactic medium (IGM) after that epoch is photoionized by the ubiquitous presence of a 
metagalactic ionizing radiation, also known as the ultra-violet background (UVB).

Usually, studies  of the so-called \lya forest (the highly ionized absorbers that give rise to the 
absorption lines in quasar spectra) assume the UVB to be spatially uniform, with the same intensity 
and spectral shape everywhere in the universe at a given redshift. 
This is a good approximation if the sources of the UVB radiation are 
uniformly distribued in space, as long as the attenuation volume is 
large enough to contain a fair sample of both the sources and the 
attenuating structures (Zuo 1992). This last condition would imply a
low cosmic optical depth, $\tau$, to ionizing photons after
reionization ($ z < z_{i}$). The cosmic optical depth should be low enough infact, to insure 
the attenuation lengh to be larger than the typical separation between
sources producing the UVB, which is $\lambda_s \sim 100 h^{-1}$ Mpc
(comoving) for quasars and more than an order of magnitude less for
galaxies. 
The relative contribution of these two populations to the UVB is unclear. Different 
independent studies found that the contribution from galaxies should be at least comparable to that 
of quasars (Giallongo \etal 1997; Shull \etal 1999; Steidel \etal 2001; Kim \etal 2001, Bianchi 
\etal 2001, Sokasian \etal 2003) in the redshift range 2-3. 
As a caveat, we recall that Meiksin (2005) has pointed out that	 
QSOs might dominate the UVB in the redshift range $2<z<3$; according 
to his study QSO counts are highly uncertain near $z=3$ and seem to be 
sufficient to account for the UVB at $z=2$.
The expected signature of a significant  
contribution from galaxies is a considerable softening of the UVB spectrum compared to previous 
models assuming quasars as the sole sources of ionizing radiation (Haardt \& Madau, 1996).

Although the IGM at $z < z_{i}$ is highly ionized, cosmic mildly overdense regions which give rise 
to the \lya forest contain enough neutral hydrogen and singly ionized helium to significantly 
attenuate the Lyman continuum flux from the sources of the UVB. As a result, the condition 
$\lambda_p \gg \lambda_s$ might not be fulfilled. For this reason, the metagalactic ionizing flux 
has been modeled starting from the intrinsic spectra of the ionizing sources, subsequently filtered 
through the IGM (Haardt \& Madau, 1996; Fardal, Giroux \& Shull, 1998). In brief, these 
calculations consist in reprocessing the UVB photons in a clumpy photoionized universe, by solving 
the radiative transfer (RT) equation in one dimension. 
The space density along the photon propagation direction, and physical properties of the sources 
(as well as of those of the absorbing clouds) are assumed as the average of the observed values, 
and the spectral redshift evolution of the UVB, assumed to be uniform in space, is derived.

Clustering of the ionizing sources and  scattering in their emission properties, together with 
inhomogeneities in the IGM density field can introduce significant fluctuations both in the intensity 
and spectral shape of the UVB.

Indeed, increasingly strong observational evidence for a significantly variable metagalactic ionizing 
radiation has been accumulating through a number of studies of the \HI and \HeII \lya forest. 
In a highly photoionized gas the ratio of \HeII to \HI column density, defined as the $\eta$ parameter,
is proportional to the ratio of \HI to \HeII photoionization rates (Fardal \etal 1998): 
$\eta~\equiv~N_\nHeII/N_\nHI\propto \Gamma_{\nHI}/\Gamma_{\nHeII}$. 

Kriss \etal (2001) and Shull \etal (2004) analyzed the FUSE observations of the fluctuating \HeII 
absorption toward the bright quasar HE 2347-4342 at $z=2.885$. Using \HI \lya data from Keck and HIRES 
(High Resolution Echelle Spectrograph), they found an \HI counterpart for more than 50\% of the \HeII 
absorption lines. The measured $\eta$ values, cover a wide range from 1 to 1000 (with an average 
$\langle \eta \rangle \approx 80$).  Analogous observations of the \HeII \Lya forest 
towards HS 1700+6416, by Remiers \etal (2004), reveal a similar $\eta$ variation. 
The large scatter is a clear signature from an ionizing background which is significantly fluctuating 
throughout the IGM. Shull \etal (2004) pointed out two important points: (i) a small scale $\eta$ 
variability (on typical scales of $\Delta z \sim 10^{-3}$, corresponding to about 1$h^{-1}$ Mpc 
comoving at $z \sim 3$), and (ii) an observed correlation of high-$\eta$ (i.e. soft ionizing spectra) 
absorbers with low density regions (voids in the \HI \lya distribution). The authors further suggest 
that these effects, confirmed by Reimers \etal (2004) observations, might be caused either by 
spatial/spectral fluctuations of the ionizing sources on scales of 1 Mpc, or by RT effects trough a 
filamentary IGM whose opacity variations control the penetration of 1-5 Ryd radiation over 30-40 Mpc 
distances (a combination of the two is also possible).

Although recent HST (Hubble Space Telescope) observations (Telfer \etal 2002) show a broad 
distribution of QSO spectral indices from $\alpha \simeq$ 0 to 3 ($J_\nu \propto \nu^{-\alpha}$ 
representing the quasar spectrum) this spread could hardly be advocated to explain the small scale 
$\eta$ fluctuations. The reason is that the QSO mean separation ($\approx 100$~Mpc comoving at 
$z \sim 3$) exceeds by far the characteristic scale of about 1$h^{-1}$Mpc on which the $\eta$ 
variations are observed. Radiative transfer through an inhomogeneous medium appears to be the most 
natural origin for such fluctuations. In turn, these might result from a number of effects, such as 
shadowing, self-shielding and filtering of the radiation.
All these effects act on the same scales of the inhomogeneities in the IGM, thus on scales smaller 
than 1 Mpc. It is then important to study the correlation function of both $J_\nu$ and $\eta$ with 
the density fluctuations.

The first attempt to account for the RT effects of a background ionizing radiation was by 
Nakamoto, Umemura \& Susa (2001). They focused on the reionization of an inhomogeneous universe 
looking at the evolving configuration of the neutral hydrogen distribution, but not directly at the 
fluctuations in the radiation field. Nonetheless, emphasizing the deviations of the neutral hydrogen 
distribution obtained in their simulation from the case of a perfectly transparent medium embedded in 
a uniform ionizing background, they indirectly stressed the importance of RT effects such as 
self-shielding and shadowing by translucent regions, which indeed have been well reproduced by 
their simulations.

Radiative transfer is furthermore expected to have some effect on the IGM temperature. 
Previous studies (Abel \& Haehnelt 1999; Bolton \etal 2004) focused on the effects of RT on the 
temperature evolution during reionization, \ie behind an ionization-front (I-front) expanding in 
a neutral or mildly ionized IGM. These analysis have shown that radiative transfer can boost the 
mean gas temperature with respect to the case in which an instantaneous, uniform reionization 
occurs (\ie its OT counterpart), as a consequence of filtering effects. 
In fact, the ionizing spectrum across the I-front becomes harder as it expands further 
from the ionizing source, because most of the low-energy photons (having higher cross-sections) are 
absorbed by recombining atoms inside the \HII region, and its photo-heating power increases 
accordingly. The overall predicted RT effects are: (i) an enhancement of the IGM temperature by a 
factor $\approx 2$ after reionization; (ii) increased scattering in temperature-density relation;  
(iii) a strong hardening of the metagalactic ionizing flux after the \HeII reionization. 
Here we consider the more stationary situation in which the mean ionization level in the simulation 
volume remains roughly constant; hence strong RT signatures on the IGM temperature are not expected. 
Nevertheless we look for a systematic dependence of the photoheating rates on the density. 

So far, much of the work in studing UVB fluctuatios have be done to
asses the effects produced by inhomogeneous distribution of ionizing 
sources, almost neglecting or just approximately including the effects
of RT through the filamentary IGM.
Zuo (1992a,1992b) first developed a theory to deal with the
fluctuations produced by randomly distributed ionizing sources, in
terms of the probability distribution of the ionizing UV radiation
field (Zuo, 1992a) and of its two-point correlation function (Zuo,
1992b).
Meiksin \& White (2003) made a step forward by appying Zuo's formalism
to ionizing sources distribution derived via numerical simulations of
structures formation.
A few groups have studied, trough numerical simulations, the
effects on the statostical properties of the \lya forest of UVB 
fluctuations induced by source clustering or non-uniform emissivity
properties as light-cone effects, QSO time variability, lifetime
and spectral index scattering. 
Gnedin \& Hamilton (2002) first investigated  via RT simulations the 
effects of UVB fluctuations induced by quasars inhomogeneous distribution, on the matter 
power-spectrum inferred from the observed/simulated trasmissivity of the \lya forest 
(Croft \etal 1999). By comparing the mock spectra from RT simulations and from simulations assuming 
photoionization equilibrium with a uniform UVB, they found the effects of RT to be negligible for 
the power spectrum; nevertheless it is interesting to notice that, in order to match the observed 
transparency of the IGM, the mean photoionization rate required when including properly RT, has to 
exceed by 20\% the one required in the OT approximation. 
Meiksin \& White (2004) attempted to quantify the UVB fluctuations and examined the possible effects
on the power spectrum and auto-correlation function of the \Lya forest. They calculate 
the UVB intensity smearing out the radiation emitted from a discrete distribution of QSOs 
in the simulated volume, within an attenuation length whose evolution with redshift is evaluated 
modeling the gas density distribution with a PM simulations. 
Their calculation separately adds the contribution of Lyman Limit System (LLS), which are not resolved 
in numerical simulations, on the basis of their observed statistical properties.
Croft (2004) uses a time-independent ray tracing algorithm to model the space density of ionizing 
radiation produced by QSOs at $z\approx 3$, neglecting its feedback on the ionization state of the IGM.
The study is based on N-body simulations and prescriptions must be used to relate the dark matter
density and the \HI optical depth. 
McDonald \etal (2004) use a rough self-shielding approximation, in which the background radiation in 
each cell is attenuated by the column density of that cell; shadowing and filtering effects are 
neglected altogether. The most relevant outcome of these studies for the present one is that UVB 
fluctuations induced by the discreteness of the UVB sources could be only relevant at high redshifts 
and on large scales. 
Below $z = 4$ the importance of such fluctuations rapidly decreases, as a consequence of the increase 
of $\lambda_p$ which smooths the flux at a given position over a larger ensemble of sources.

For this reasons we take here a complementary, and still unexplored, approach which consists in the 
investigation of UVB fluctuations introduced by the RT effects mentioned above. 
We focus on possible detectable signatures of these effects on quantities derived from \lya forest 
spectra,as photoionization rate fluctuations, the temperature-density relation 
and $\eta$ parameter distributions. To this aim we have performed RT simulations, applying \CR~ 
(Maselli, Ferrara \& Ciardi, 2003, hereafter MFC03) to a cosmological density field  
exposed to an ionizing background radiation characterized by a spatially constant emissivity. 
Such radiation field, by interacting with the filaments of the cosmic web, is rapidly turned into a 
fluctuating one. The rest of the paper is devoted to study and quantify this effect.

\section{Numerical simulations}

The aim of the present study is to estimate the effects of RT on the metagalactic ionizing radiation 
and the resulting effects induced in the physical properties of the IGM. 
The present study is focused at $z\approx 3$. We have exposed a snapshot at 
a fixed redshift of a cosmological density field to an (initially) uniform
ionizing radiation field. 
The density field has been derived with the numerical code described in Marri \& White (2002), a 
modified, multiphase version of the entropy-conserving SPH code GADGET2 (Springel \etal 2001). 
Several algorithms have been designed to  reduce artifacts which occur in cold, dense gas clouds 
embedded in a hot diffuse halo; the code includes a new implementation of feedback which allows 
supernova energy to be channeled effectively into the heating of diffuse gas and the evaporation 
of cold clouds. 
A multiphase description of the interstellar medium is adopted, based on an explicit separation of 
protogalactic gas into diffuse and dense (star forming) components which is able to suppress star 
formation, reheat cold gas and drive outflows from galactic disks. The presence of a UV background 
produced by QSOs  is included in the OT approximation following Haardt \& Madau (1996) from which the 
IGM temperature and ionization state are calculated assuming photoionization equilibrium.

We use here a snapshot at $z=3.27$ from a simulation which has been used and analyzed in previous 
works of our group (Bruscoli \etal 2003; Maselli \etal 2003; Fangano, Ferrara \& Richter 2005, 
in preparation). These studies have shown that the statistical properties of the IGM, e.g. the 
PDF of the transmitted \lya flux, are correctly reproduced. The simulation uses $128^3$ particles 
in a 10.5$h^{-1}$ comoving Mpc box (corresponding to a mass resolution  of $8.5 \times 10^6M_\odot$ 
in gas) and assumes a $\Lambda$CDM cosmological model with $\Omega_0=0.3$, $\Omega_\Lambda=0.7$, 
$\Omega_b=0.04$, $h=0.7$ and $\sigma_8=0.9$; periodic boundary conditions are assumed.

\subsection{Radiative Transfer}

We have performed the RT calculation using the code \CR. A detailed description of the code, 
its performances and several test problems can be found in MFC03, to which we defer the interested 
reader. 
In the present study we neglect 
the effects of cosmological expantion; this choise is motivated by the fact 
that the light crossing time of an attenuation leght ($\sim 100$ Mpc) is 
roughly one tenth of the Hubble time at the redshift of interest.
We have exposed the density field obtained from the SPH simulation to a UVB radiation with 
spatially constant emissivity.  
The UVB is subsequently modified as it propagates in  the filamentary 
density field by self-shielding, shadowing and spectral filtering effects, finally resulting in a 
fluctuating field. 
The RT simulation has been evolved for a physical time, $t_s=0.7 \times 10^7 {\rm yr}$. 
This choice has been made, after some experiments, to guarantee that numerical convergence of the
solution is achieved. 
It has to be noticed that here the simulation time does not correspond to a physical evolution of
the background, but it is the time required for the gas ionization and temperature to reach 
equilibrium in the self-consistently radiation field shaped by radiative transfer. 
In the following we discuss two crucial, albeit somewhat technical, points concerning a new 
implementation of a uniform radiation background in \CR~ and its flux calibration within the
simulation.

\subsubsection{Simulating a uniform UVB}

\begin{figure}
\centerline{\psfig{figure=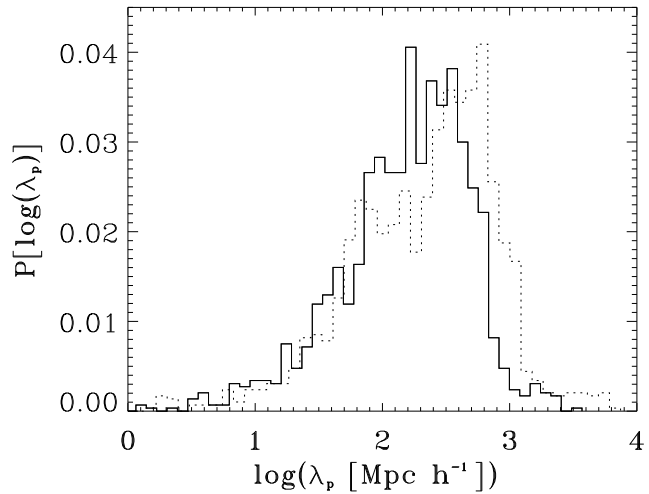,height=6.5cm}}
\caption{Distribution of the comoving mean free path of ionizing photons at $z=3.27$ 
as derived from the SPH simulation. The solid curve includes opacity of both \HI and \HeII; 
dotted curve is for \HI only.}
\label{fig01}
\end{figure}

The algorithm described in MFC03 for the implementation of the presence of a uniform background 
radiation in the simulation box has been improved in several ways. In the original implementation, 
background radiation was modeled by allowing box boundary cells to emit photons packets with the 
prescribed UVB spectral energy distribution. Although this choice is suitable for some type of 
problems it cannot match the high accuracy required by the present study. 
To better fulfill such requirement we have devised a new implementation 
in which all cells in the box, but the ones with overdensity 
$\Delta =\rho/\langle\rho\rangle \le \Delta_e$, behave as UVB 
radiation sources, in the sense that they emit photon
packets\footnote{It is important to notice that here the cells
emitting photons are not meant to reproduce physical sources of UV 
radiation; the adopted method aims at reproducing
a uniform UVB in the simulated volume, within a Monte Carlo ray
tracing scheme.}.
The threshold on the overdensity is meant to preserve the effect of 
self-shielding in high density regions and galactic halos. 
For definiteness, we take the value corresponding to the typical overdensity of collapsed halos at 
the virial radius. The exact value of $\Delta_e$ depends on the density profile of the halo; it can 
be shown that it is $\approx 59.2$ for a isothermal profile and it is $\approx 63.7$ for the NFW 
profile (Navarro, Frenk \& White 1996). In this paper, $\Delta_e = 60$ has been assumed. We adopt 
periodic boundary conditions, so that packets can travel trough the simulation box for a number of 
cycles, $N_{cyc}$, given by the ratio between the mean free path of ionizing photons, $\lambda_p$, 
and the physical size, $L$, of the simulated volume. 
We have derived an estimate for $\lambda_p$, running a benchmark 
simulation aimed at evaluating the statistical distribution of the mean
free paths of photons -whose frequencies are distributed according to
the HM96 spectrum- in the assigned density field. 
For this purpose, in the benchmark simulation we just follow the 
propagation of photons packets, i.e. not cosidering their effect on the gas 
properties; due to the statistical treatment of the packet energy 
deposition inherent to the Monte Carlo scheme (see MFC03), we define as ``absorbed'' a packet whose 
initial energy has been depleted by $> 95\%$. In Fig. \ref{fig01} we show the mean free path
distribution obtained from such numerical experiment. The solid curve corresponds to the mean free 
path obtained by including the opacity of both hydrogen and helium absorbers, whereas the dotted curve
 corresponds to \HI opacity only. 
The mean (median) values of the two distributions are 300 and 495 (199 and 306) $h^{-1}$ Mpc comoving.
This shows that the \HeII contribution to the diffuse IGM opacity is significant. 
The above value of $\lambda_p$,  for  \HI opacity, is consistent with that inferred from Fan \etal (2002), 
$\lambda_{p} \approx 100 h^{-1}$ Mpc at $z \approx 3$, by comparing observations of high resolution quasar 
spectra with semianalytical models of the \lya forest with a simple schematic model of the IGM opacity based 
on a peak statistics approach. 

Guided by this finding, and given our box size, we fix $ N_{cyc} = 10$.  The agreement of the 
simulated and observed mean values of $\lambda_p$ further supports the 
$N_{cyc}$  choice above. Periodic boundary conditions allow us to simulate volumes smaller than 
the mean free path of neutral hydrogen ionizing photons, having thus a higher spatial
resolution, without much information loss. However, we make clear that the main aim of the present 
study are RT effects on scales of $\approx 1$~Mpc, the characteristic scale 
of IGM filaments; no attempt is made to describe large scale fluctuations induced by non-uniform 
distribution of the UVB sources.
%
%

According to the \CR~ Monte Carlo algorithm tracking photon packet propagation, the mean number 
of crossings events per cell is given by \beq \label{ncross}\langle{\cal
N}_{cross}\rangle=\frac{N_{cyc} N_p}{N_c^2} \eeq where $N_p$ is the total number of photon packets 
in a single run, and $N_c^3$ the number of grid cells in the 
computational volume\footnote{In the OT case, a single packet crosses typically $N_{cyc}N_c$ cells, 
thus the total number of crossings in a simulation is $N_p N_{cyc} N_c$ for $N_c^3$ cells 
(for more details see MFC03)}. The new background implementation has been tested by running a
simulation in which we force the condition $\tau=0$. 
Although, when not inlcuding cosmological expansion, the $\tau=0$ 
assumption would imply an infinite radiation filed (\ie the Olber's
paradox), we get around this problem by fixing $N_{cyc}$ to a finite
value. All the parameters then, are setted in order to reproduce the
required UVB intensity, as described in Sec. 2.1.1.
Ideally, in this case  the intensity of the
UVB should be perfectly uniform, as fluctuations caused by photon-matter interactions (absorption 
and scattering), redistributing the ionizing radiation in space and frequency,
are suppressed. In practice, uniformity can be achieved only up to a certain accuracy level, 
due to finite number packet statistics, which produces noise fluctuations in the {\it emission}
pattern. The amplitude of  such fluctuations decreases with increasing $N_p$, which is however 
constrained by computational expense. We have performed different test runs 
increasing $N_p$ to accurately quantify the amplitude of such fluctuations.  The computed r.m.s. 
of the fluctuations, $\sigma_N$, around the mean value of the UVB intensity 
as a function of the numerical parameters of the run are shown in Table 1. The value for $N_c$ and 
$N_{cyc}$ are the same adopted for the actual cosmological simulations\footnote{The intrinsic scatter 
$\sigma_N$, associated to the numerical resolution of a given run, does not depend on the physical 
time of the simulation, $t_s$, and on the intensity of the UVB which, as discussed later, depends 
on the normalization factor $A$.}. These fluctuations are due to the photon shot noise intrinsic to 
the adopted numerical algorithm. 

From Tab. 1, we see that runs at higher resolution (i.e. higher values of $N_p$) are characterized 
by a small intrinsic scatter of
the radiation intensity: $\sigma_N$, decreases from 20\%  
for $N_p=3 \times 10^5$, to less than 1\% for $N_p=3 \times 10^8$.
At the same time, we have also checked that during these test runs $\langle {\cal N}_{cross}\rangle$ 
closely matches the theoretical 
value (as expressed in eq.~\ref{ncross}). 
As a good compromise between accuracy and computational time we will adopt $N_p=3 \times 10^7$ for the 
cosmological RT simulations. 
This value, according to the above tests, guarantees that the ionizing radiation field fluctuations 
are uniformly distributed within a precision level of 4\%.
Hence, larger fluctuations seen in the simulation outputs can be safely attributed to RT effects.
\begin{table*}
\centering
\caption{Parameters of the simulation test runs used to check uniformity and calibration of the 
radiation field.}
\begin{tabular}{||c|c|c|c|c|c||}
\hline\hline
  & & Uniformity  Tests &  & & \\ \hline 
$A$ & $N_p$ & $N_{cyc}$ & $N_c$ & $t_s [yr]$ & $\sigma_N$  \\ \hline 
$-$ & $3 \times 10^5$ & 10 & 128  & $-$ & 20 \% \\ \hline 
$-$ & $3 \times 10^6$ & 10 & 128  & $-$ & 10 \% \\ \hline 
$-$ & $3 \times 10^7$ & 10 & 128  & $-$ &  4\% \\ \hline 
$-$ & $3 \times 10^8$ & 10 & 128  & $-$ &  1\% \\ \hline 
\hline
                                      &             & Calibration  Tests &  & & \\ \hline 
$0.95 \times 10^{52}$ & $10^6$ & 10 & 32  & $10^5$ & 1\% \\ \hline 
$0.95 \times 10^{52}$ & $10^6$ &  5 & 32 & $10^6$ &  10\% \\ \hline 
$0.95 \times 10^{52}$ & $10^7$ & 10 & 64 & $10^7$ & 3\% \\ \hline 
$0.95 \times 10^{52}$ & $3 \times 10^8$ & 10 & 128 & $10^7$ & 2\% \\ 
\hline
\label{table1}
\end{tabular}
\end{table*}
%
%
\begin{figure*}
\centerline{\psfig{figure=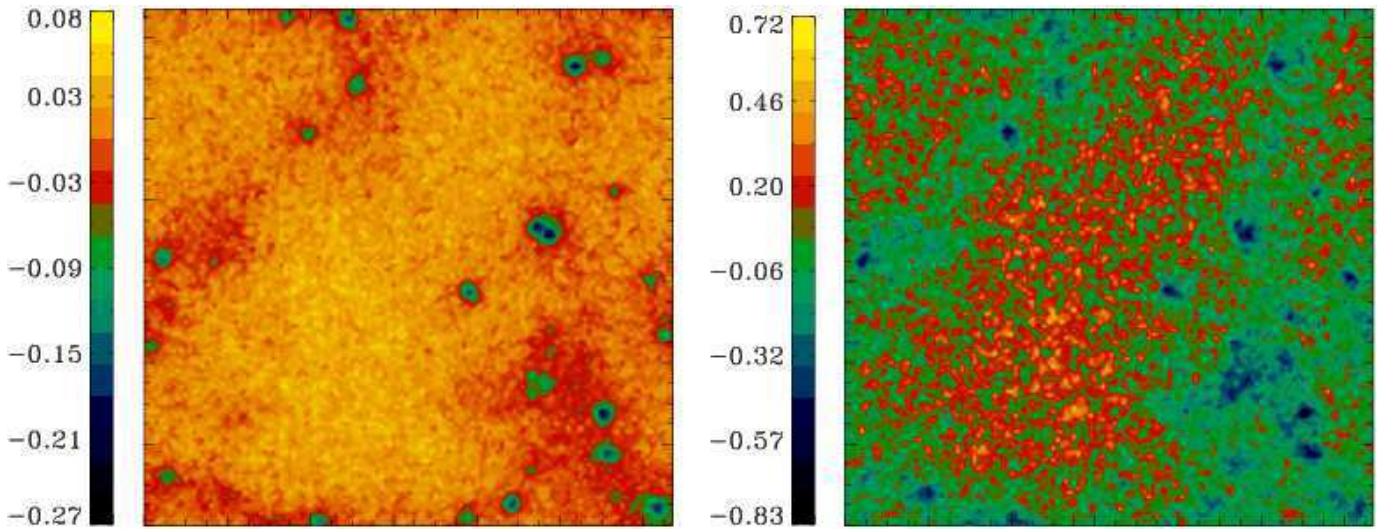,height=7.cm}}
\caption{Spatial fluctuations of  \HI  (left panel) and \HeII (right) photoionization rates in a 
slice across the RT simulation 
of (physical) depth $\Delta x \approx 27$ kpc.
The curves denote the mean values.}
\label{map2D}
\end{figure*}
\subsubsection{UVB intensity calibration}
Particular attention must be devoted to the ionizing flux calibration. The problem arises when the 
emissivity has to be related to the photon content of packets originating 
from a cell, as it is not obvious how convert the latter into the specific flux units in which  
$J_\nu$ is usually assigned (erg~s$^{-1}$~cm$^{-2}$~Hz$^{-1}$~sr$^{-1}$ ).  
To overcome the problem we have devised a calibration technique which turned out to be highly precise 
and 
straightforward.

Given a set of simulation parameters, $N_p$, $N_{cyc}$, $N_c$ and $t_s$, the intensity of the 
radiation 
field can be tuned by varying the number of (monochromatic) photons contained in a packet, $N_\gamma$, 
at emission. We express this quantity as follows
\beq 
\label{ngamma}
N_\gamma= A  \;\; \Delta t \;\;  \frac{J(\nu)}{ J_{912}}
= A \;\; \frac{N_c^2\; t_s}{N_p\; \; N_{cyc}} \frac{J(\nu)}{ J_{912}}
\eeq
where $A$ is the calibration factor\footnote{Note that the calibration factor $A$ has physical 
dimensions s$^{-1}$, \ie a photoionization rate.} 
to be determined and $\Delta t$ is the mean time interval between 
two subsequent crossings of a cell by photon packets; for the second equality see MFC03.
A nice feature of the previous expression is its scaling with the simulation parameters. 
Once the intensity and the UVB spectral shape are fixed, the factor $A$ is the same for any 
combination of the numerical parameters.
An order of  magnitude for $A$ can be derived by equating the theoretical and numerical estimates of 
the ionization fraction increment, $\Delta x_{\nHII}$, for a cell during the time interval $\Delta t$. 
We will consider here, for the sake of clarity, the case of a pure hydrogen gas and a monochromatic 
spectrum 
($J(\nu)/J_{912}=1$). The theoretical value is given by 
\beq
\label{Th}
\Delta x_{\nHII}^{t}=\Gamma_{\nHI} x_{\nHI} \Delta t;
\eeq 
similarly,  the numerical one can be expressed as
\beq
\label{Num}
\Delta x_{\nHII}^{n}=N_\gamma (1-e^{-\tau_c}) /N_{\nHI}= A \Delta t\tau_c /N_{\nHI}
\eeq 
where $\tau_c$ is the cell optical depth, $N_{\nHI}=n_{\nHI} \Delta x^3$, with $\Delta x^3$ being the 
cell 
volume, is the total number of H-atoms in the cell; the last equality holds for the OT case 
($\tau_c \ll 1$).
By equating the two expressions, one can write $A$ as
\beq
\label{AA}
A= \frac{\Gamma_{\nHI}}{\sigma_0} \Delta x^2,
\eeq
where we have used $\tau_c\approx~n_{\nHI} \sigma_0 \Delta x$, with $\sigma_0$ being the \HI 
photo-ionization 
cross-section at the Lyman limit. 
For the RT simulations used in this study, $\Delta x \simeq 27$ kpc; hence for a typical 
$\Gamma_{\nHI}=10^{-12}$~s$^{-1}$,  $A  \approx 10^{51}$~s$^{-1}$. 

The determination of the exact value of $A$ for the general case of a H/He gas and of a 
non-monochromatic
spectrum, is not as straightforward.
The adopted strategy has been to select a cell at the center of the simulation box, to assign it  
physical
properties characteristic of the hydrodynamic simulation (we choose $n=10^{-5}$cm$^{-3}$, $T=10^4$K, 
$x_\nHI$ the value at the 
photo-ionization equilibrium with the assigned $\Gamma_\nHI$ value). We assume $\tau =0$ for all 
remaining cells in the box; 
this guarantees that the ionizing radiation field impinging on the target cell 
has not been modified either in intensity or spectral shape by RT effects. 
In the above configuration, we run \CR to  compute the \HI fraction in the central cell 
and the average photoionization rate, $\Gamma_\nHI$, by counting the cumulative number of ionizing 
photons absorbed in a given simulation time. We then tune the value of $A$ by requiring that the 
$x_\nHI$ and 
$\Gamma_\nHI$ estimated values match the assigned values within an accuracy of 1\%.
We find that $\Gamma_{\nHI}=1.2 \times 10^{-12}$~s$^{-1}$ (HM96) corresponds to 
$A=1.55 \times 10^{52}$s$^{-1}$.\\
As discussed above, the same value of $A$ reproduces correctly the assigned ionization rate
(for a given UVB spectral shape) independently of the values of $N_p$, $N_{cyc}$, $N_c$ and $t_s$,
provided that $N_\gamma$ is evaluated according to eq. \ref{ngamma}. 
The degree of accuracy is shown in the bottom part of Tab 1 where the r.m.s. deviation, $\sigma_N$, 
of the distribution of  $\Delta x_{\nHII}^{n}$ with respect to the theoretical value 
$\Delta x_{\nHII}^{t}$ is shown for different combinations of the numerical parameters. 
The scatter is always less than 10\%, depending on the resolution of the run. 
The calibration method adopted results thus to be accurate and stable. \\
In actual RT simulations, we should further account for the opacity of the IGM in order to reproduce 
at each cell the correct ionizing flux\footnote{With ``correct'' we mean the ionizing flux which 
reproduces the assigned mean opacity.}; the photon content of packets is infact depleted 
while it propagates trough the box, due to non-zero opacity of the IGM distribution. To counteract 
this effect, we boost the normalization parameter $A$ by 2.3; the exact value of this factor has been 
determined empirically, requiring that the mean opacity of the simulated volume is the same for the 
OT (zero-opacity) and RT cases. If the calibration factors for the RT and OT cases are related 
according to the equation $A^{OT}=A^{RT}e^{-\tau_{eff}}$, it corresponds to an effective 
optical depth $\tau_{\rm{eff}}\approx 0.8$.

Extracting the correct physical values of the photoionization and photoheating rates (at each cell) 
from the RT simulation is not straightforward as well. Again, we need a reference OT simulation 
(with the same numerical parameters) in order to calibrate correctly these values. The OT and RT 
simulations differ (i) in the normalization factor, $A^{OT}$ and $A^{RT}$ related as above, and (ii) 
in the fact that for the OT we do not allow for absorption to occur. 
In both RT and OT cases, summing up the contribution given at each crossing event, the following 
relations hold:
\be
\label{rates}
\Gamma_i&=&  C_i \sum_{j\in cross}N_\gamma(\nu_j)\sigma_i(\nu_j), \;\;\;\;\;\;\;\;\;\;\;\; \\ 
{\cal H}_i&=& C_i^\prime  \sum_{j\in cross}N_\gamma(\nu_j)\sigma_i(\nu_j)(\nu_j-\nu_{th,i}), \nonumber 
\ee
where $j$ is the index counting the crossing events in a given cell and $\nu_j$ is the frequency 
of the packet at the $j$-th crossing; $C_i$ and $C_i^\prime$ are a scaling factor which depend only 
by the numerical parameters and are the same for both the OT and RT cases. The other symbols follow 
the conventional notation.   
Having determined $A^{OT}$ as seen above, we can safely impose that $\langle\Gamma_i^{OT}\rangle=\Gamma_I^{th}$ 
(\eg $\Gamma_H^{OT}=1.2\times10^{-12}$s$^{-1}$) and accordingly derive $C$, which in turns 
is used to derive the correct $\Gamma_i^{RT}$ physical values at each cell in the box.

Finally, we mention that in the RT simulation the emission of diffuse radiation from recombining gas 
(estimated to account for about 30\% of the ionizing background at $z=3$) has been turned off inside 
{\tt CRASH}. This is done to prevent double-counting of such radiation, which is already included in 
the Haardt \& Madau (1996) computation of the adopted UVB. 

\section{Results}

In this section we present the results of the  RT simulations. 
We first analyze the amplitude of RT-induced fluctuations as traced by UVB photo-ionization 
and photoheating rates of the three absorbers \HI, \HeI and \HeII. 
These are defined as 
\be
\delta \Gamma_{i}=\frac{\Gamma_i-\langle\Gamma_i\rangle}{\langle\Gamma_i\rangle},
~~~~~ \delta {\cal H}_{i}=\frac{{\cal H}_i-\langle{\cal H}_i\rangle}{\langle{\cal H}_i\rangle};
\ee
the index $i\in\{\HI,\HeI,\HeII\}$ denotes the absorbing species and 
the averages are done over all the cells at the end of the RT simulation
We then focus on the effects of such fluctuations on the IGM $\eta~\equiv~N_\nHeII/N_\nHI$ 
distribution, the $\eta$-density relation and on the IGM temperature-density relation, 
with the aim of identifying possible detectable signatures. 

\subsection{UVB fluctuations}

Spatial  fluctuations of the \HI and \HeII  photoionization rates are shown in Fig. \ref{map2D} 
on a slice across the RT simulation box of (physical) depth $\Delta x \approx 27$ kpc, 
corresponding to one grid cell ($\delta \Gamma_{\nHI}$ and $\delta \Gamma_{\nHeII}$ in the left and 
right panels respectively). RT effects result in significantly fluctuating photo-ionization 
rates on scales of few comoving Mpc, whose structure is closely linked to the underlying density 
fluctuations. As shown in the following, typically regions of low overdensity 
(say $\Delta \la 1$, often referred to as voids) have $\delta\Gamma > 0$, 
whereas overdense regions tend to have negative fluctuations. 
Comparing the two maps one can also notice that $\Gamma_{\nHeII}$ fluctuations are much stronger than 
those of $\Gamma_{\nHI}$. In general, though, RT induces fluctuations of several tens of percent of 
the mean ionizing flux in both species. Such deviations from a homogeneous radiation field are 
produced by the combined action of shielding and shadowing effects. Although these features can be 
qualitatively understood from simple physical arguments, their quantitative evaluation can only be 
done through full 3D RT simulations. \\
%
%
\begin{figure*}
\centerline{\psfig{figure=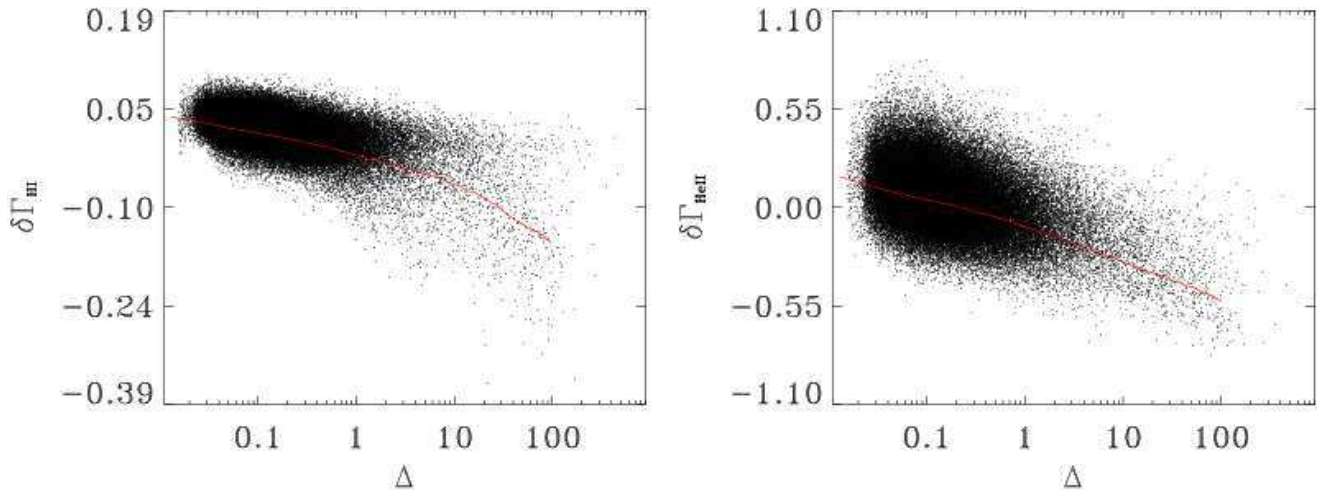,width=17.4cm}}
\caption{Spatial fluctuations of  \HI  (left panel) and \HeII (right) photoionization rates,  
as a function of the local overdensity, $\Delta$. Each point in the figures corresponds to a cell of 
the simulated grid; the solid lines show the mean $\delta \Gamma$ value as a function of $\Delta$.}
\label{gammas}
\end{figure*}
In order to make the above statements more quantitative, we show in Fig. \ref{gammas} the 
$\delta \Gamma_{\nHI}$  and $\delta \Gamma_{\nHeII}$ distributions as a function of the local 
overdensity, $\Delta$. Each point in the plots corresponds to a cell of the simulated grid. 
The plot in left panel shows clearly how the mean amplitude of \HI photoionization rate fluctuations 
varies with overdensity, passing from positive values (of the order of a few percent) in the voids 
to negative ones, larger than 10\%,  
towards higher densities. $\delta \Gamma_{\nHeII}$, shown in the right panel, exhibits an analogous 
trend, although with larger excursion, with values $\approx +25$\% ($\approx -50$\%) 
at the lowest (highest) density boundaries of the density range. 
Deviations from the mean value can be as high as 20\% and 60\%, for $\Gamma_{\nHI}$ and 
$\Gamma_{\nHeII}$ respectively, in structures corresponding to overdensities with $\Delta\gtrsim 50$,  
where therefore the UVB intensity appears to be significantly suppressed.
The steeper dependence and larger scatter of $\delta\Gamma_\nHeII$--$\Delta$ relation reflect the 
fact that the universe is more opaque in \HeII than in \HI. 
Note also that in underdense regions $\Gamma_{\nHeII}$ can exceed by more than 40\% 
the volume averaged mean value, $\langle\Gamma_{\nHeII}\rangle$.\\
We focus now on mildly overdense regions, such as filaments, which give raise to absorption lines in 
the \lya forest. These regions are exposed to roughly the mean photoionization rates for \HI, 
but the photoionization of \HeII is here suppressed by about 20\% with respect to the mean value. 
This fact suggests that the \HeIII fraction, which both in numerical simulations and data 
interpretation of \Lya forest is usually derived 
assuming photoionization equilibrium with a uniform UVB , is probably seriously overestimated.
%
%
%
\begin{figure*}
\centerline{\psfig{figure=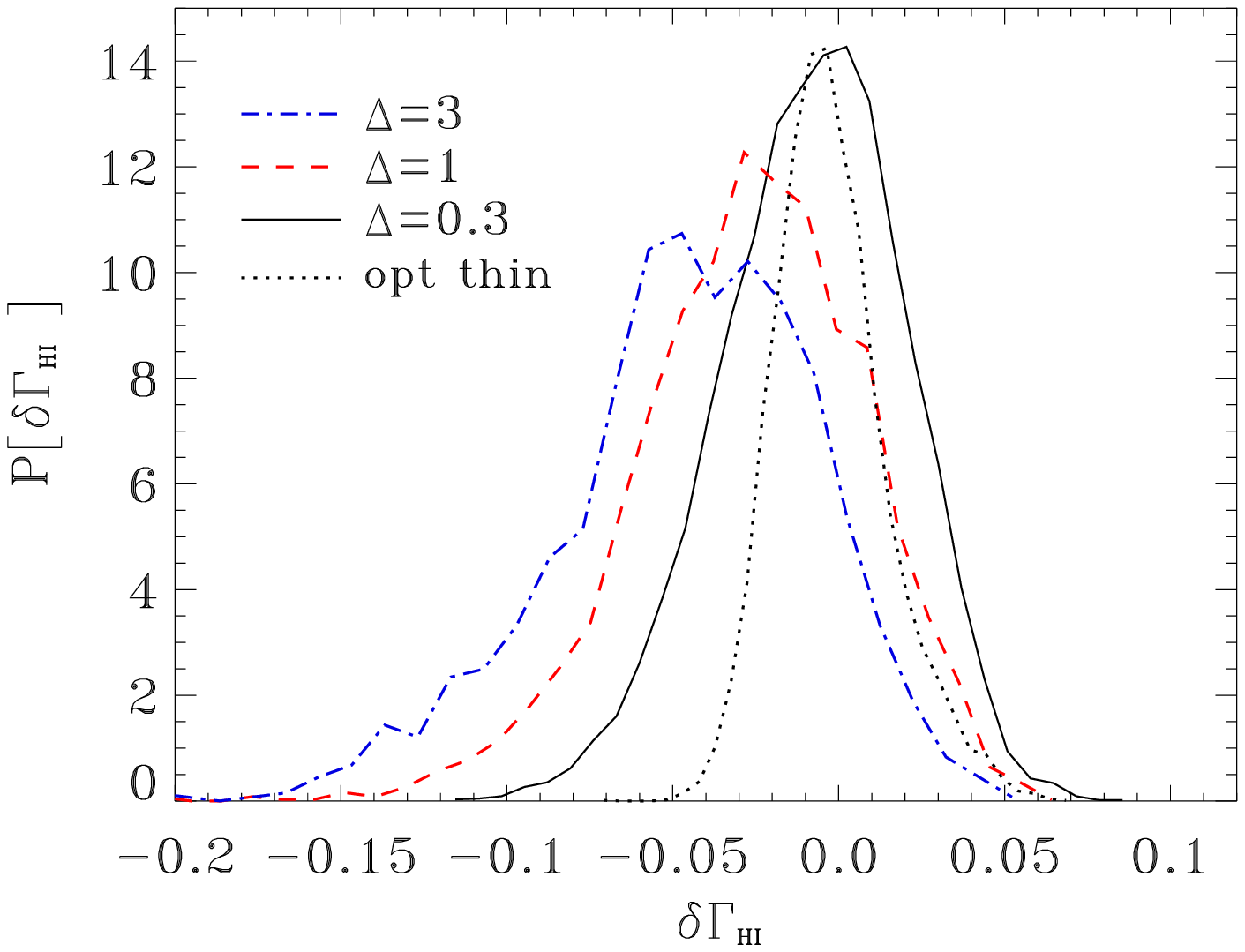,height=6.5cm}\hspace{0.2cm}
\psfig{figure=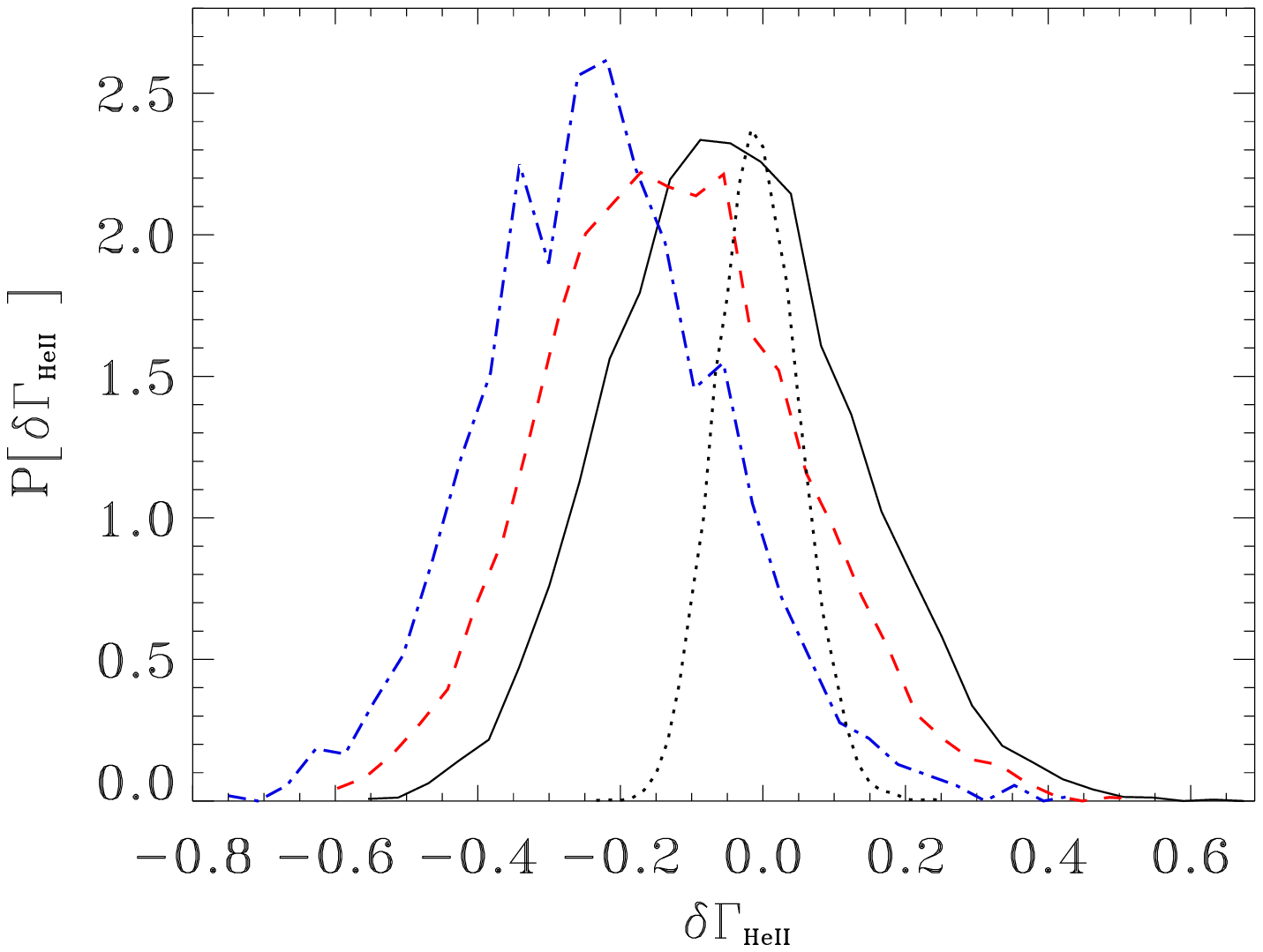,height=6.5cm}}
\caption{Probability distribution function of the fluctuations for three selected values of the 
overdensity, $\Delta=0.3$ (solid line), $\Delta=1$ (dashed)  and $\Delta=3$ (dotted-dashed). The left
and right panels refer to $\delta \Gamma_{\nHI}(\Delta)$ and $\delta \Gamma_{\nHeII}(\Delta)$, 
respectively. The dotted curves are the analogous distributions at $\Delta=0.3$ for the OT case.}
\label{cuts}
\end{figure*}
The ultimate physical nature of the decreasing trend of $\delta\Gamma_i$, with $i\in\{$\HI,\HeII$\}$, 
with density is a mixture of self-shielding and shadowing effects. 

\begin{figure}
\centerline{\psfig{figure=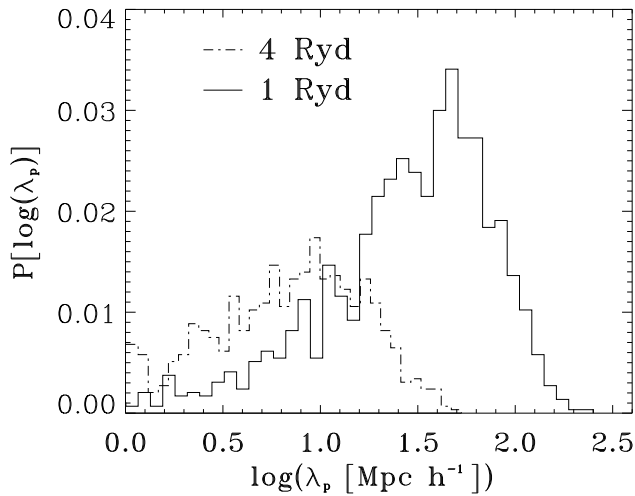,height=6.5cm}}
\caption{Probability distribution of the mean free path 
of photons at 1 Ryd (solid curve) and 4 Ryd (dotted-dashed curve).}
\label{tau_RT}
\end{figure}

It is very instructive to inspect the probability distribution function of the fluctuations at fixed 
overdensities. Fig. \ref{cuts} shows such distribution at $\Delta\in\{0.3, 1, 3\}$ (solid, dashed and 
dotted-dashed curves respectively), for  $\delta \Gamma_{\nHI}(\Delta)$ (left panel) and 
$\delta \Gamma_{\nHeII}(\Delta)$ (right panel). The distributions resemble Gaussian curves whose 
mean value decreases with $\Delta$. The shift of the mean is caused by the increasing importance of 
the shielding and it is about five times larger for \HeII ($= - 0.25$) than for \HI ($= -0.05$): 
as explained above, the attenuation of the flux due to the structures shielding is more prominent 
above 4 Ryd due to the larger opacity at those energies. In the case of \HI, a significant increase 
of the distribution width is observed towards larger overdensitites; contrarily, the three 
distributions obtained for $\delta\Gamma_\nHeII$ show a similar dispersion. This results from the 
fact that, while the \HI optical depth of the IGM is dominated by rare, overdense regions, the \HeII 
optical depth is non-negligible even in voids. This fact is confirmed if one look at the probability 
distribution for the mean free path of photons at 1 Ryd and 4 Ryd, shown in 
Fig. \ref{tau_RT} (solid and dotted-dashed curves respectively).
For 1 Ryd photons the mean free path distribution is highly peaked at roughly 60 $h^{-1}$ Mpc 
\footnote{Note that these values are lower than the ones in Fig. \ref{fig01}, because here we 
consider only photons at the \HI and \HeII ionization threshold.}, 
showing that these photons are preferentially absorbed in high 
density peaks. 
The 4 Ryd photons mean free path distribution is wider and has a mean value of about 7 $h^{-1}$ 
Mpc, which is a clear indication for a significant contribution of voids to the \HeII opacity.

We have checked that the dispersion of the flux distribution shown in Fig. \ref{cuts} do not result 
from spurious effects introduced by the emission algorithm, 
comparing the distributions above with the analogous ones derived in the OT 
simulation\footnote{We use the OT simulation we have run to test the accuracy of the uniformity of the 
emission algorithm, with $N_p=3 \times 10^7$, having thus the same numerical resolution of the RT 
simulation.}. The latter are shown in the panels of Fig.    \ref{cuts} by the dotted curves for 
$\Delta=0.3$ (as the OT distribution does not depend on overdensity, we only plot this case). 
For visualization purposes, these curves have been normalized so that their peaks match the peaks 
of the corresponding RT distributions. 
The dispersion of these distributions is obviously unphysical as it represents the intrinsic error of 
the numerical algorithm, which is of the the order of 4\% 
for \HI, as already pointed out in Table 1, and more than twice for \HeII ($\sim 10$\%); 
the higher numerical dispersion for \HeII is due to the worse spectral sampling of energies above 4 Ryd
\footnote{The number of photon packets emitted with energy above the \HeII threshold is roughly one 
order of magnitude smaller less than $N_p$.}. 
Both for \HI and \HeII, the dispersion induced by RT in the photoionization rates values is 
significantly larger than the one originating purely from numerical noise and, in the case of \HI, 
it is increasingly so at larger values of $\Delta$. Hence, this confirm that the effects we are 
finding are of physical origin and the corresponding fluctuations are a truly genuine by-product 
of UVB filtering through cosmic density structures. \\                                        

\begin{figure}
\centerline{\psfig{figure=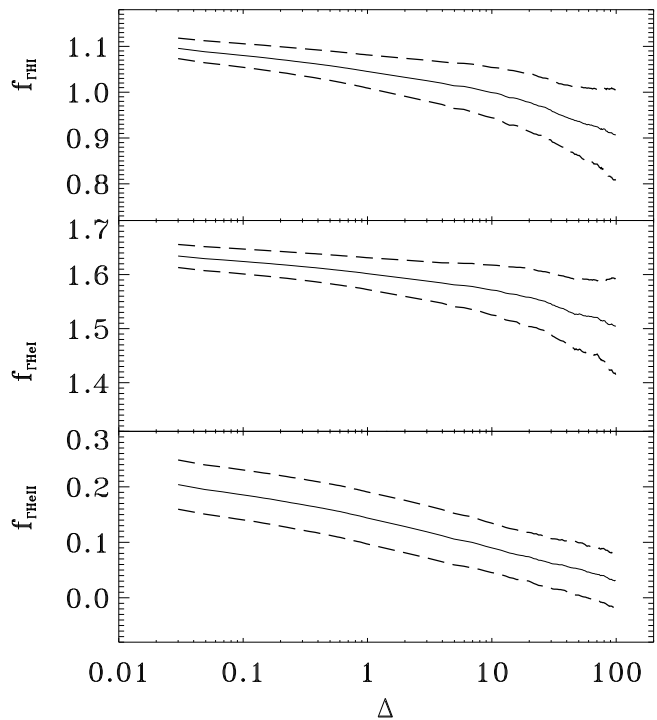,width=8.6cm}}
\caption{Photo-ionization rate correction factors (see eq. \ref{cfact}) for \HI (upper panel), \HeI 
(central) and \HeII (bottom), applicable at $z\approx 3$. The three panels show both the mean value 
(solid line) and the 1$\sigma$ statistical error in each density bin.}
\label{gammas_CF}
\end{figure}

We have found so far that RT effects induce significant fluctuations in the metagalactic ionizing 
radiation; these could be effective in deviating the actual IGM ionization state from the one 
inferred assuming a uniform ubiquitous UVB and hence should be accounted for in high-precision 
studies of the \lya forest. However, performing full RT simulations is extremely expensive and 
prohibitive in maximum-likelihood analysis.
The tight correlation between fluctuations in the photo-ionization rate and 
overdensity can be used to {\it correct} the commonly adopted mean values derived 
in calculations (e.g. Haardt \& Madau 1996; Fardal \etal 1998) carried out with 1D RT calculations, 
under the assumption of photo-ionization equilibrium with a uniform UVB. \\
In order to cast our results in a form which could be useful for future studies, we have computed the 
photo-ionization rate correction factors from our RT simulations. These are defined as 
\beq 
\Gamma_i^{RT}(\Delta)=f_{\Gamma_i}(\Delta)~\Gamma_i^{OT}(\Delta);
\label{cfact}
\eeq
their mean values are plotted in Fig. \ref{gammas_CF} for \HI, \HeI and \HeII (upper, central and bottom 
panel respectively) along with the associated 1$\sigma$ errors. 
These results are valid for $z \approx 3$. 
In a companion paper we will study the evolution with redshift of the UVB fluctuations.
RT transfer effects are generally more important in overdense regions, although the correction 
factors derived in the present study can be reliably used for $\Delta<100$, a limit dictated both 
by mass resolution of the SPH simulation and shot noise associated to the limited number of photon
packets. Differently from the case of \HI, for which $\langle f_{\Gamma_{\nHI}}\rangle \approx 0.98$ 
is close to unity, the mean value of the correction factors for \HeI and \HeII are 
$\langle f_{\Gamma_\nHeI}\rangle \approx 1.56 $ and  $\langle f_{\Gamma_{\nHeII}}\rangle \approx 0.2$
(averages are evaluated over the overdensity range $0.03 < \Delta < 100$) . 
This reflects mean opacities differences among the three absorbers. In the OT approximation, 
the IGM opacity is higher in \HI than in \HeI and, as a consequence, the UVB increase necessary to match 
the IGM \HI \lya opacity in RT simulations results in a higher mean $\Gamma_\nHeI$. For the same 
reason, as the IGM OT opacity is by far higher in \HeII than in \HI,  $\langle\Gamma_{\nHeII}\rangle$ 
is highly suppressed with respect to the OT value when RT effects are properly treated.
Again, our results demonstrate that the contribution to the \HeII opacity of the diffuse IGM 
(as opposed to discrete absorbing systems) is relevant. 
The same arguments explain the steeper dependence of the \HeII correction factor from density and its 
larger dispersion. 

A few remarks are worth concerning the correction factors. First, we remind that the adopted flux 
calibration  (see Sec. 2) guarantees that the mean \HI opacity in the RT simulation box corresponds 
to that at the photoionization equilibrium with the uniform UVB in the
OT case. We found that the mean value of $\Gamma_{\nHI}$ in the two
cases (RT and OT) are the same within few percent. 
Gnedin \& Hamilton (2002) found a different result: in order to reproduce the same \HI \lya IGM 
trasmissivity in their RT calculation, they require a 20\% higher volume-averaged \HI 
photoionization rate with respect to the OT case.
The reasons for the discrepancy are not clear, but 
it is probably due to the fact that while Gnedin \& Hamilton (2002) account for RT during the 
reionization process (\ie ionizing a neutral gas), here we re-distribute the ionizing radiation 
accounting for RT effects in a pre-ionized gas.  
Meiksin \& White (2003) also found that fluctuations required a
boost in the radiation field,
comparable to Gnedin \& Hamilton; however, in a more recent
investigation, they found that the boost is
reduced to few percents at $z$\ltsima4  when correlations in the
radiation field due to RT were included (Meiksin \& White 2004). This
seems to suggest that the boost depends on the magnitude of the correlations
in the radiation field.\\
It is furthermore interesting to note that $\Gamma_\nHI$ can be fitted with a power-law of the form: 
$\Gamma_\nHI(\Delta)= \Gamma_\nHI^0 \Delta^{- a_1}$ with $a_1\simeq 0.022$. 
In the calculation of optical depth to the \lya scattering given by (\eg Weinberg \etal 1999; 
McDonald \& Miralda-Escud\'e 2001)
\beq
\tau=\tau_0\frac{\left(1+z\right)^6\left(\Omega_bh^2\right)^2}{T^{0.7}H(z)\Gamma_{-12,\nHI}(z)}\Delta^\beta,
\eeq
it is thus possible to account for UVB spatial fluctuations adopting a {\it corrected} index 
$\beta^\prime=\beta+a_1$, \ie $\tau\propto \Delta^{\beta^\prime}/ \Gamma_{-12,\nHI}^0$, 
\ie in a variation of the {\it apparent} IGM equation of state.
Even though at $z\approx3$ we found a negligible variation of $\beta$ (only few percents), 
we expect the above effect to be important at higher redshifts.
%
%
%
%
%
\begin{figure}
\centerline{\psfig{figure=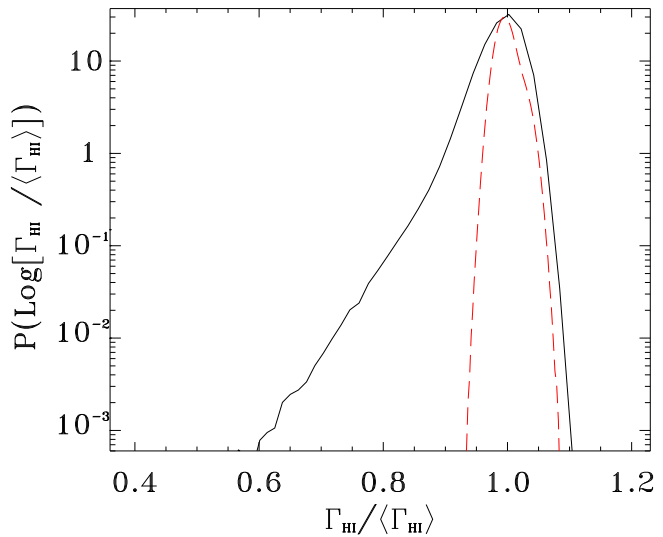,width=8.6cm}}
\caption{Comparison between the PDFs of the hydrogen photo-ionization rates,  normalized with respect 
to the mean value, from full RT simulations (solid line), and OT ones (dashed).}
\label{flux_PDF}
\end{figure}

Fig.   \ref{flux_PDF} shows the PDF of the hydrogen photo-ionization rates resulting from our 
simulations (normalized to the volume averaged mean value $\langle\Gamma_\nHI\rangle$): 
the solid curve refers to the full RT simulation, whereas the dashed one has been obtained from the 
OT simulation. The RT distribution is much wider than the OT one, showing an extended tail towards
lower values of $\Gamma_\nHI/\langle\Gamma_\nHI \rangle$ which is caused by shadowing and shielding 
due to gaseous structures. There is also a slight increase of cells seeing an ionizing flux above the 
mean, occurring as a result of the flux amplitude redistribution. 
Further interesting aspects can be inferred comparing the UVB fluctuations produced by RT effects 
and those induced by the inhomogeneous distribution of rare quasars, assumed to be the dominant UVB 
sources, derived by Croft (2004) and shown in Fig.   5 of that paper. 
Even in that case, the PDF of the ionizing flux appears to be asymmetric: however, the asymmetry is 
in the opposite direction. 
A similar result is obtained by Meiksin \& White (2003) as
shown in Fig.4 of that paper, where the effects of the sources
stochastic distribution in space are studied.
In fact, fluctuations originating from sources inhomogeneous distribution are characterized by 
considerable power in the high flux tail ($\gtrsim 1.5$ times the mean value), corresponding to the 
enhanced ionizing flux in the vicinity of the luminous ionizing sources. 
Hence, both RT and inhomogeneous distribution of ionizing sources,
contribute to broaden the radiation field distribution, basically increasing 
the power on opposite sides of the flux mean value.   
It is important to realize though, that the spatial character of these fluctuations is very different. 
The signature of source inhomogeneities is seen on scales comparable to the typical separation of the 
sources, i.e. $> 100$~Mpc for quasars. RT-induced fluctuations are instead found 
on a few (comoving) Mpc scales.
%
%
\begin{figure}
\centerline{\psfig{figure=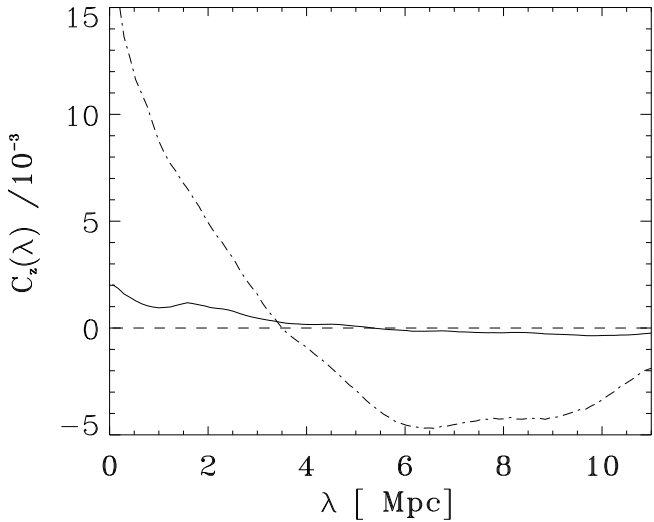,width=8.6cm}}
\caption{Correlation function (comoving scale) of photoionization rates for \HI (solid line) and 
\HeII (dotted-dashed line). }
\label{corr_func}
\end{figure}
To clarify this point we have studied the spatial correlation of RT-induced UVB fluctuations, 
computing the $\delta \Gamma_i$ correlation function by using the following formula:
\beq
{\cal C}_i(\lambda) =\sum_{j=1}^{N_c}\sum_{k\ge j}^{N_c}{\frac{\delta \Gamma_i(j)~\delta \Gamma_i(k)}
{N_\lambda}}.
\eeq
Here the index $j$ and $k$ span all the cells in the simulation grid, $\lambda$ is the comoving 
separation between the $j$-th and $k$-th cells and $N_\lambda$ the number of pair cells whose 
distance is the range $[\lambda, \lambda+\Delta \lambda]$. 
We have similarly calculated a second correlation function, ${\cal C}_i^r(\lambda)$, obtained from a 
random realization of $\delta\Gamma_i$, in the same volume and with the same mean as 
${\cal C}_i(\lambda)$.  The difference, 
$\bar{\cal C}_i(\lambda) = {\cal C}_i(\lambda) - {\cal C}_i^r(\lambda)$ is plotted in 
Fig. \ref{corr_func} for \HI and \HeII.  The zero-point of these functions gives the characteristic 
correlation comoving length of the RT induced fluctuations, which is found to be roughly 
5.5 and 3.5 Mpc comoving for \HI and \HeII respectively. 
Although this results should be checked for convergency, by performing
analogous simulations in larger volumes with the same resolution
(Meiksin \& White 2004), they seem to be plausible. The correlation
scales we have found, could be associated with the typical 
size/separation of filaments (i.e. mildly overdense regions) of the
IGM which are clearly identified in the maps presented in 
Fig.  \ref{map2D}. These structures have optical depths $\gtrsim 0.1$ both in hydrogen
and helium,  and therefore can leave their imprint in the fluctuation field.\\
Meiksin \& White (2003,2004) have shown that inhomogeneous sources
(assumed to be QSOs) distribution can produce correlations which are
significant even on small scales. Both the RT induced fluctuations and the
ones originating from disomogeneous sources distribution, should then
be treated at once, in order to derive a correlation function which is
reliable for the study of the pixel fluxes statistic in high-$z$ QSOs
spectra.

\subsection{\HeII/\HI ratio fluctuations}
\begin{figure}
\centerline{\psfig{figure=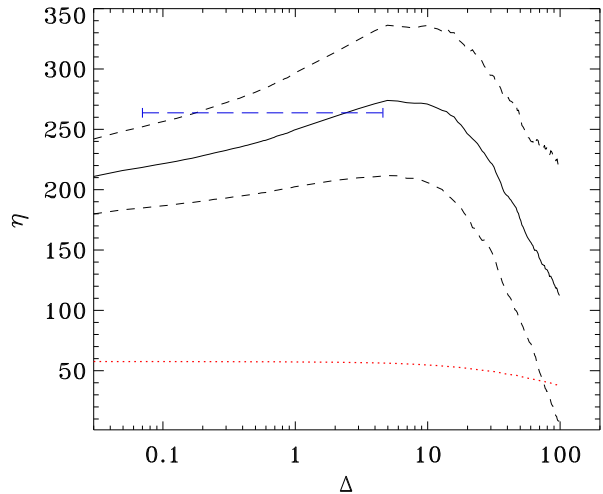,height=6.5cm}}
\caption{The parameter $\eta$ is shown as a function of the overdensity $\Delta$: the solid (dotted) 
line represents the mean $\eta$ value from the RT (OT) simulation and the dashed curves correspond to 
1$\sigma$ r.m.s. deviation of the RT distribution. 
The long-dashed line shows the mean $\eta$ value from Reimers et. al. (2004) in the range of 
the observed densities (we exlcude data points with  
$\Delta > 7$, as at higher densities the uncertainties on the $\eta$ values 
inferred from observations become very high and metal line contamination is a problem).}
\label{eta_fig}
\end{figure}

The $\eta$ parameter, defined as the ratio of \HeII to \HI column density of an absorber in the \lya 
forest, gives informations on the intensity and on the spectral properties of the metagalactic 
ionizing radiation. Assuming photoionization equilibrium in highly ionized H/He gas, the following 
expression can be derived (Fardal \etal 1998):
\beq
\label{eta_eq}
\eta \equiv \frac{N_{\rm HeII}}{N_{\rm HI}}=\frac{\alpha^{(A)}_{\rm HeII}}
{\alpha^{(A)}_{\rm HI}}\frac{n_{\rm HeIII}}{n_{\rm HII}}\frac{\Gamma_{\rm HI}}{\Gamma_{\rm HeII}}.
\eeq
Here $\alpha^{(A)}_{\rm HI}=2.51 \times 10^{-13}T_{4.3}^{-0.76}$ cm$^3$ s$^{-1}$ and 
$\alpha^{(A)}_{\rm HeII}=1.36 
\times 10^{-12}T_{4.3}^{-0.70}$ cm$^3$ s$^{-1}$ are the case $A$ recombination rate coefficients, 
appropriate in the low density regime. 
We have derived the $\eta$ parameter, for each cell of the numerical grid, from  eq. \ref{eta_eq} 
with $n_{\rm HeIII}/n_{\rm HII}$ and $\Gamma_{\rm HI}/\Gamma_{\rm HeII}$ taken from the RT simulation.
The results are plotted in Fig. \ref{eta_fig}: the solid (dotted) line represents the mean $\eta$ value
derived in the RT (OT) simulation and the dashed curves are the 1$\sigma$ r.m.s. deviations of the 
distribution. 

In the OT case, as expected , $\eta$ shows  little dependence on $\Delta$ as a result of the 
fact that the three ratios in eq. \ref{eta_eq} are roughly constant. The ratio between the 
recombination coefficients slowly evolves with temperature, $\propto T_{4.3}^{0.06}$; the  
$\Gamma_\nHI/\Gamma_\nHeII$ ratio does not vary by construction if the UVB is spatially uniform;
finally for $\Delta < 10$, the gas is fully ionized  thus 
$n_{\rm HeIII}/n_{\rm HII}\approx n_{\rm He}/n_{\rm H}=0.785$ is roughly constant.   
At higher densities  $n_{\rm HeIII}/n_{\nHII}$ at the photoionization equilibrium starts to decrease 
because \HeII recombines faster than \HII and $\Gamma_{\rm HeII}/n_{\rm He}$ is roughly one order 
of magnitude lower than $\Gamma_{\rm HI}/n_{\rm H}$. 
The mean $\eta$ value from our OT simulation is $\eta\approx50$, consistent with the expected 
(from eq. \ref{eta_eq}) value, $\eta \sim 0.43 \times \Gamma_{\rm HI}/\Gamma_{\rm HeII} \approx 52$.

RT effects are crucial to detexrmine the actual values of  $\eta$ in the IGM. This can be seen 
by comparing the previous results for the OT case with the ones from full RT calculations,
shown in the same figure; the two cases differ from each other both in the mean value and in shape. 
The RT mean $\eta$ value is larger than the OT one by about a factor of 
four because of the different amount of attenuation suffered by \HI and \HeII ionizing radiation. 
As we have seen in Sec. 3, the \HeII ionizing flux is strongly suppressed 
with respect to the OT case, because even the most underdense voids in photoionization equilibrium 
with the UVB contains enough \HeII to remove photons with energies above 4 Ryd from the background 
radiation; the higher values we find for $\eta$ in the RT simulation can be interpreted in terms of 
the ratios between the correction factors discussed in Sec 3. The $\eta$ parameter shows furthermore 
a strong dependence on overdensity: $\eta$ slowly grows with density, reaches a maximum around 
$\Delta \approx 7$ and then rapidly declines  to values  which are still  well in excess of the 
OT ones. In fact, at low density $n_{\rm HeIII}/n_{\nHII}$ is roughly constant, and density dependence
is governed by $\Gamma_{\rm HI}/\Gamma_{\rm HeII}$. This ratio increases with density as the universe 
is more opaque in \HeII than in \HI, as already discussed. When the density is high enough, 
$n_{\rm HeIII}/n_{\nHII}$ decreases more rapidly than the growth of 
$\Gamma_{\rm HI}/\Gamma_{\rm HeII}$, due both to the latter effect and to \HeIII recombining faster 
than \HII. This results in the inverted trend of the $\eta$ dependence on $\Delta$. 

We compare our results from RT simulations with observational data by Reimers \etal (2004) in 
Fig. \ref{eta_fig}, using the relation $\Delta=0.8 N^{0.7}_{HI,13}$ (Madau, Ferrara \& Rees 2001). 
As the data released are still preliminary and could suffer from errors associated with the analysis 
technique (Fechner \etal, 2004; private communication), we here consider the mean $\eta$ value for 
$\Delta < 7$, as at higher densities the uncertainties on the $\eta$ values inferred from observations
become very high and metal line contamination is a problem. 
Although it is important to point out that the dispersion in the data is substantial.\\
The comparison in Fig. \ref{eta_fig} clearly shows that the mean $\eta$ value from observations  
is inconsistent with the OT results, \ie the expected value for the case of a spatially uniform UVB.
The higher $\eta$ measured value can be explained when accounting for the radiative transfer of the 
metagalactic ionizing radiation. 
Furthermore at $\Delta>7$, the RT simulation output predicts an $\eta$--$\Delta$ 
anti-correlation. This suggests that RT effects, such as shielding and 
shadowing, can explain the systematically lower $\eta$ values observed in mildly/highly overdense
absorption systems. 
According to our results, the lower values of $\eta$ associated with high density absorption systems 
do not arise, as previously thought, from the hardening of UVB spectrum produced by the presence of 
local hard-spectrum sources. Instead, we find  that the metagalactic ionizing flux is softer in 
filaments than in voids. Our preferred interpretation of the $\eta$ trend seen in Fig. \ref{eta_fig} 
is that the $\eta$--$\Delta$ anti-correlation is a direct  consequence of the fact that in high 
density regions $n_{\rm HeII}$ increases much faster than $n_{\rm HI}$.

%
%

\subsection{Effects on the IGM Temperature}
%
%
\begin{figure}
\centerline{\psfig{figure=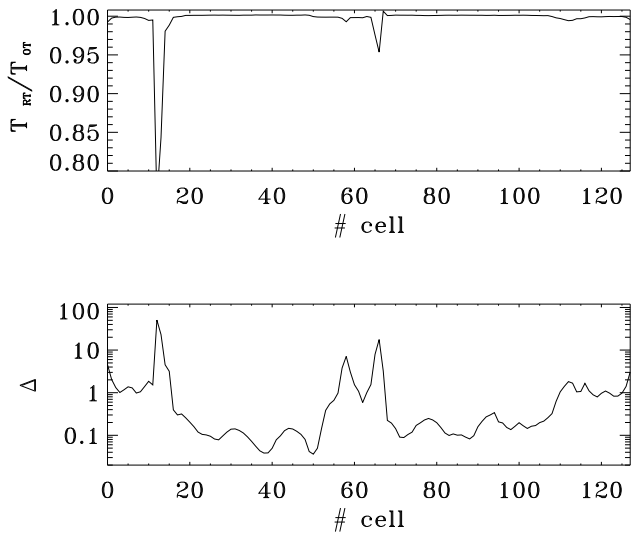,height=7.cm}}
\caption{{\it Upper panel:} Ratio between the temperature in the RT  simulations, $T_{RT}$, and the 
temperature estimated from the SPH simulation in the OT limit, $T_{OT}$ along a typical light of sight.
{\it Lower panel:} The overdensity, $\Delta$, along the same line of sight.}
\label{temp1}
\end{figure}

Fluctuations in the photoionization rates could in principle affect the 
IGM temperature. In this paragraph we attempt to quantify such effect.  The upper panel in 
Fig. \ref{temp1} shows the ratio between the temperature in the RT  simulations, $T_{RT}$, 
and the initial temperature, $T_{OT}$ (estimated in the SPH simulation in the OT limit), along a 
random line of sight (los); the lower panel shows the overdensity, $\Delta$, along the same los.  
The temperature ratio $T_{RT}/T_{OT}$ is significantly different from unity only in regions 
corresponding to high density peaks, where the temperature shows deviations of the order of  $1-2$\%. 
The temperature in low density regions is basically unchanged. This is not at odd with previous 
studies (Abel \& Haehnelt 1999; Bolton \etal 2004) on the effect of radiative transfer on the 
temperature of the ionized gas, which indeed suggested that RT can be effective in boosting the 
IGM temperature by at least factor of two during the reionization process or in general when 
considering the photo-ionization of a neutral gas, with respect to the case in which 
a the gas is instantaneously photoionized and photo-heated by a uniform UVB. 
As discussed in the Introduction, such boost in temperature is due to the filtering of the 
ionizing radiation. The latter is significantly harder at the edge of the I-front, getting harder 
as the I-front moves further from the source; this is due to the fact that the large cross-section, 
low-energy photons are selectively absorbed inside the HII region where they compensate for recombinations. 
The harder spectrum at the I-front yields a higher heating power with respect to analogous one 
provided by the unfiltered spectrum and, due to the long characteristic cooling times, the 
low density IGM retains memory of the initial higher temperature phase. 
In the present study we focus on the transfer of ionizing radiation in a {\it ionized} IGM, thus 
the above filtering process is not efficient due to the low IGM opacity. 
In order to account properly for RT effects on the IGM temperature, a full and consistent simulation 
of the reionization process including RT is necessary.

\section{Conclusions}

In this study we have analyzed the effects of radiative transfer on the
spatial and spectral properties of the ionizing metagalactic radiation
at a fixed redshift, $z=3.27$. So far, this is the first attempt to study the properties
of the ultraviolet background radiation including full 3D RT calculations.
Such investigation is motivated by the increasing number of observational evidences for a
fluctuating UVB (Shull \etal 2004; Reimers \etal 2004). These
observations reveal significant fluctuations on scales of few comoving Mpc, which could hardly be
originated from the inhomogeneous spatial distribution of ionizing sources, whose mean
separation is $\approx 100$ $h^{-1}$ Mpc for QSOs at $z\approx 3$.

We here neglect the UVB inhomogeneities arising from source
clustering or non-uniform emissivity properties (which have been shown to be relevant
only on large scales and at high redshift, \ie $z>4$), and focus on smaller scales.
Our approach consists in performing  RT simulations, applying \CR to a cosmological
density field exposed to an ionizing background characterized by a spatially constant emissivity.
Such radiation field, by interacting with the filaments of the cosmic web, is rapidly turned into a
fluctuating one. \\
We found that RT effects result in relevant
fluctuations in the photo-ionization rates, 
whose structure is closely linked to the underlying density
fluctuations. The UVB mean intensity, and thus its ionizing power,
is progressively suppressed moving towards higher densities; 
such effect is found to be more important for radiation above
4 Ryd, due the high \HeII opacity of the universe.
The ultimate physical nature of the decreasing trend of photoionization rates 
with density is a mixture of self-shielding and shadowing effects. \\
Interestingly, we found that the $\delta\Gamma_\nHI$ dispersion increases towards 
higher $\Delta$ (from 8\% at $\Delta=0.3$ to roughly 15\% at $\Delta=3$), 
which is not the case for $\delta\Gamma_\nHe$. The differences in \HI and \HeII photo-ionization
rates originate  from the fact that while the \HI optical depth of
the IGM is dominated by rare, overdense regions, the \HeII optical depth is non-negligible 
even in voids.\\
We further found that shielding in overdense regions
($\Delta \gtrsim 7$) from \HeII ionizing radiation, produces a
decreasing trend of the $\eta$ parameter with overdensity; 
this trend has already emerged from
observational data (Shull \etal 2004; Reimers \etal 2004).
According to our results, the lower values of $\eta$ associated
with high density absorption systems do not arise, as previously
thought, from the hardening of UVB spectrum produced by the
presence of local hard-spectrum sources. Instead, RT simulations suggest 
that the $\eta$--$\Delta$ anti-correlation is a direct consequence of the
fact that in high density regions $n_{\rm HeII}$ increases much
faster than $n_{\rm HI}$. 
A stronger support for the presence of RT induced UVB fluctuations comes 
from the comparison of the mean $\eta$ value from Reimers \etal (2004) observations 
and our numerical results. We find in fact, that the mean observed $\eta$ value is too high
to be consistent with a uniform UVB and that, accounting properly for the filtering of 
the ionizing radiation trough the translucent IGM, it is possible to reconcile the 
observational data with theoretical expectations.

The results of the present study have several 
implications. In fact, the wealth of information derived from the
\Lya forest has been so far extracted under the assumption of a
uniform UVB, both from real and simulated data; the current
results should thus be revised in order to account for the more
realistic scenario of a fluctuating UVB. We give some examples.
The matter power-spectrum inferred from the observed/simulated
trasmissivity of the \lya forest (Croft \etal 1999) provides direct
information on the primordial universe. UVB
fluctuations could affect the power-spectrum
shape derived in previous studies. Gnedin \&
Hamilton (2002) used RT simulations to study the effects of UVB
fluctuations induced by quasars inhomogeneous distribution on the
matter power-spectrum  (however, they did not look directly at the
RT effects on the spatial UVB fluctuations), not
finding a significant variation. However, the question of whether
small scale fluctuations 
in the UVB could be relevant in determining the shape of the matter 
power-spectrum needs further study.

The UVB fluctuations we have found could provide an alternative
explanation, other than \HeII reionization occurring at $z \approx
3$, for the observed fluctuations in the $\eta$ parameter as well
as for the ones in the abundance ratio \CIV/\SiIV. As
discussed in the Introduction, it has been shown that current data are not 
inconsistent with a scenario in which \HeII is
completely reionized as early as  $z\sim6$  by thermal radiation
produced in collapsing structures (Miniati \etal 2004); the
observed fluctuating \CIV/\SiIV abundance ratio and the \HeII
patchy opacity could be then explained as direct consequences of 
UVB fluctuations arising from radiative transfer of the
metagalactic ionizing radiation trough the filamentary IGM.

Although in the present study we have carried out our analysis at $z\approx 3$, 
we can infer that the UVB fluctuations become more important
towards higher redshift due to the increasing mean opacity of the
IGM. In a future work we plan to investigate accurately the redshift 
evolution of UVB fluctuations.

Finally, it is worth reminding  that we have neglected the possible contribution 
of local sources embedded in high density regions to the ionizing radiation. 
The relative contribution of local sources embedded in quasar absorption line systems,
which are commonly identified as high density peaks in the IGM, 
has been subject of recent debate.
Schaye (2004) argued that contribution from local source dominates the 
UV metagalactic radiation in these absorption system, while 
Miralda-Escud\`e (2004) has come to the opposite conclusion, deriving
an upper limit to the importance of local sources by a consequence 
of the surface brightness conservation. 
The problem seems still unsettled  and it would be desiderable 
to further investigate it via numerical RT simulations including 
both the UVB and the local sources  radiation.\\
\\

\section*{Acknowledgments}
We acknowledge T. Choudhury \& F. Haardt for enlightening discussions, 
C. Fechner, J.Tumlinson and M. Shull for providing observational data.
This work was partially supported by the Research and Training Network
`The Physics of the Intergalactic Medium' set up by the European Community under the contract 
HPRN-CT2000-00126 RG29185. The work has been completed at the Kavli Institute 
for Theoretical Physics Program ``Galaxy-IGM Interactions'', whose hospitality is warmly acknowledged.

\label{lastpage}
\end{document}